\documentclass[multphys,vecphys]{svmult}

\usepackage{makeidx}         % allows index generation
\usepackage{graphicx}        % standard LaTeX graphics tool
\usepackage{multicol}        % used for the two-column index
\usepackage[bottom]{footmisc}% places footnotes at page bottom

\makeindex

\begin{document}

\title*{Metallicity structure in X-ray bright galaxy groups}
\author{Jesper Rasmussen \and Trevor J.~Ponman}
\institute{School of Physics and Astronomy, University of Birmingham, 
Edgbaston, Birmingham B15 2TT, UK
\texttt{jesper@star.sr.bham.ac.uk}}

\maketitle

\begin{abstract}
  Using \emph{Chandra} X-ray data of a sample of 15 X-ray bright
  galaxy groups, we present preliminary results of a coherent study of
  the radial distribution of metal abundances in the hot gas in
  groups.  The iron content in group outskirts is found to be lower
  than in clusters by a factor of $\sim 2$, despite showing mean
  levels in the central regions comparable to those of clusters. The
  abundance profiles are used to constrain the contribution from
  supernovae type Ia and II to the chemical enrichment and thermal
  energy of the intragroup medium at different group radii. The
  results suggest a scenario in which a substantial fraction of the
  chemical enrichment of groups took place in filaments prior to group
  collapse.
\end{abstract}

\section{Introduction}\label{rasm:sec1}

Hot X-ray emitting gas constitutes the dominant baryonic component in
groups and clusters of galaxies.  The chemical enrichment of this
intracluster medium (ICM) is believed to originate mainly in material
ejected from group and cluster galaxies by supernovae, with a smaller
portion driven out by galaxies through galaxy--galaxy and galaxy--ICM
interactions.  The spatial distribution of metals reflect the action
of such processes, providing insight into the mechanisms, other than
gravity, that have shaped the thermodynamic properties of gas in
groups and clusters.

Iron and silicon comprise the major diagnostic elements for studies of
the abundance distribution in groups and clusters, as these elements
give rise to the most prominent lines in the soft X-ray spectrum of
thermal plasmas, and are among the elements for which there is a
distinctively different yield in type~Ia and type~II supernova
explosions according to standard supernova (SN) models.  Whilst iron
is predominantly produced by SN~Ia, silicon is more evenly mixed
between the two SN types.  The ratio of Si to Fe abundance therefore
provides valuable information on the relative importance of SN~II vs.\
SN~Ia in enriching the intragroup medium.

Abundance profiles of massive clusters have received considerable
attention. Due to their lower X-ray luminosities, the situation in
lower-mass systems is much less clear, despite their relatively larger
importance for the cosmic baryon and metal budgets.  With the current
generation of X-ray telescopes, however, the ability to perform
spatially resolved X-ray spectroscopy has improved dramatically,
enabling detailed mapping of the metal distribution also in groups.
Here we utilize this to derive radial profiles of the abundance of Fe
and Si from \emph{Chandra} observations of a sample of 15 groups.  The
sample size allows us to explore statistical trends within the sample
and obtain a clearer picture of the content and behaviour of metals in
group outskirts, where most of the intragroup gas resides but where
the low X-ray surface brightness often prohibits robust constraints to
be obtained for individual systems.

\section{Sample and Analysis}\label{rasm:sec2}

Our sample is based on the 25 GEMS groups that show group-scale
extended X-ray emission (the 'G' sample of \cite{rasm:osmo04}). Only
groups with \emph{Chandra} archival data featuring more than 6,000 net
counts and being more distant than 20\,Mpc were selected, while at the
same time discarding obviously unrelaxed groups. We added four more
groups conforming to these criteria, leaving a sample of 15 reasonably
relaxed groups with high-quality \emph{Chandra} data.

All data sets were subjected to standard screening criteria and
cleaned for periods of high background, with blank-sky background data
filtered in an identical manner. Background-subtracted spectra were
extracted in annuli and fitted with a {\sc vapec} model in {\sc xspec}
using Solar abundances from \cite{rasm:grev98}. All elements were tied
together, except for Fe and Si.  To put all groups on an equal
footing, all radii were normalized to $r_{500}$, the radius enclosing
a mean density of 500~times the critical density, using an empirical
relation based on the temperature of the group \cite{rasm:fino01}. The
temperature profiles themselves reveal that most of the groups feature
a central region of cool gas, referred to here as the group core.  For
each group, Table~\ref{rasm:tab1} lists $r_{500}$ and the mean
temperature and abundance derived outside the cool core in the radial
range 0.1--0.3\,$r_{500}$.

\begin{table}
\centering
\caption{Basic X-ray properties of the group sample, with mean temperatures 
  $\langle T \rangle$ in keV, abundances $\langle Z \rangle$ in Z$_\odot$, 
  and $r_{500}$ in kpc}
\label{rasm:tab1}       
\begin{tabular}{lcccclcccc}
\hline\noalign{\smallskip}
Group & $\langle T \rangle$ & $\langle Z \rangle$ & $r_{500}$ & \hspace{3mm} &
Group & $\langle T \rangle$ & $\langle Z \rangle$ & $r_{500}$ \\
\noalign{\smallskip}\hline\noalign{\smallskip}
{\bf NGC~383}  & 1.65$^{+0.04}_{-0.06}$ & $0.39^{+0.07}_{-0.07}$ & 578 & &
{\bf NGC~4125} & $0.33^{+0.12}_{-0.05}$ & $0.19^{+0.17}_{-0.07}$ & 259 \\
{\bf NGC~507}  & $1.30^{+0.03}_{-0.03}$ & $0.40^{+0.07}_{-0.06}$ & 513 & & 
{\bf NGC~4325} & $0.99^{+0.02}_{-0.02}$ & $0.39^{+0.08}_{-0.06}$ & 448 \\
{\bf NGC~533}  & $1.22^{+0.05}_{-0.05}$ & $0.28^{+0.07}_{-0.06}$ & 497 & &
{\bf HCG~62}   & $1.00^{+0.03}_{-0.03}$ & $0.12^{+0.02}_{-0.03}$ & 450 \\
{\bf NGC~741}  & $1.42^{+0.14}_{-0.12}$ & $0.16^{+0.08}_{-0.06}$ & 536 & &
{\bf NGC~5044} & $1.12^{+0.03}_{-0.03}$ & $0.25^{+0.04}_{-0.03}$ & 476 \\
{\bf NGC~1407} & $1.01^{+0.07}_{-0.09}$ & $0.15^{+0.09}_{-0.06}$ & 452 & &
{\bf NGC~5846} & $0.66^{+0.04}_{-0.03}$ & $0.16^{+0.06}_{-0.05}$ & 366 \\
{\bf NGC~2300} & $0.78^{+0.04}_{-0.03}$ & $0.17^{+0.07}_{-0.05}$ & 397 & &
{\bf NGC~6338} & $2.13^{+0.19}_{-0.07}$ & $0.25^{+0.08}_{-0.05}$ & 657 \\
{\bf HCG~42}   & $0.80^{+0.05}_{-0.05}$ & $0.25^{+0.31}_{-0.10}$ & 402 & & 
{\bf NGC~7619} & $1.06^{+0.07}_{-0.03}$ & $0.23^{+0.05}_{-0.05}$ & 463 \\
{\bf NGC~4073} & $1.78^{+0.07}_{-0.09}$ & $0.49^{+0.07}_{-0.07}$ & 600 \\
\noalign{\smallskip}\hline
\end{tabular}
\end{table}

\section{Fe and Si Radial Profiles}\label{rasm:sec3}

\begin{figure}
\centering
\includegraphics[height=4.8cm]{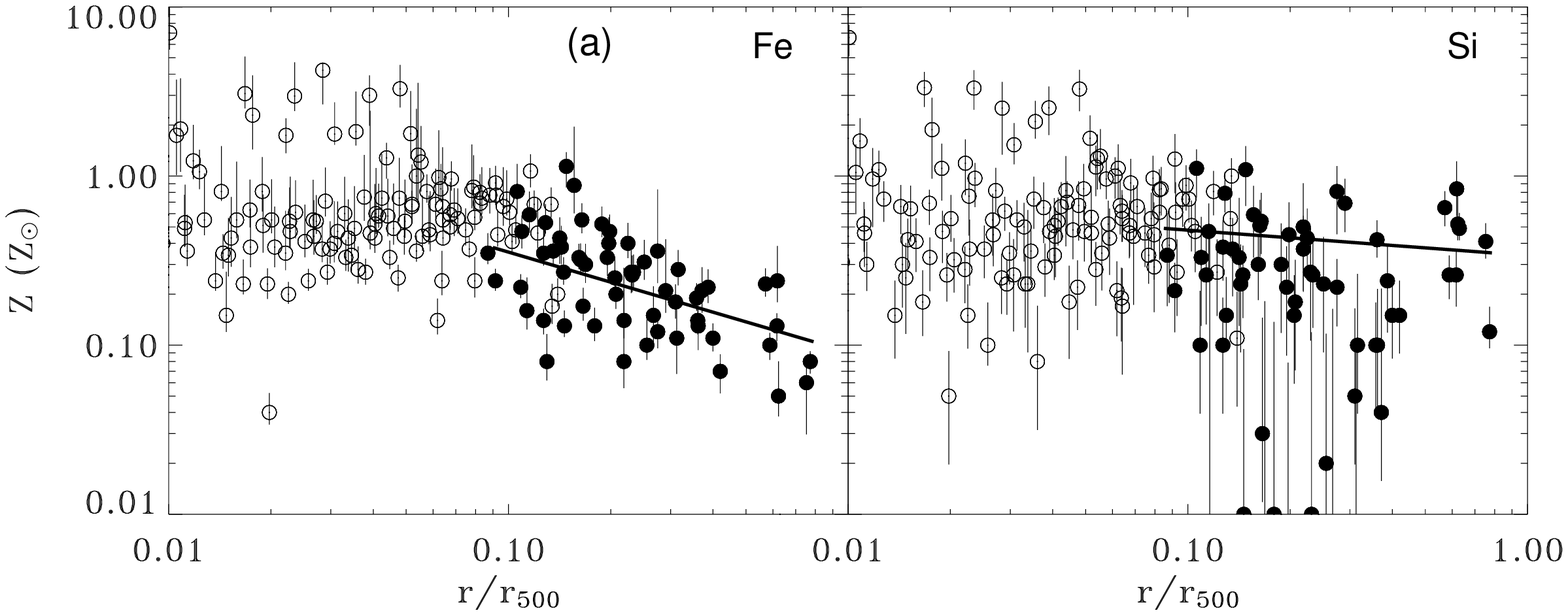}
\includegraphics[height=4.8cm]{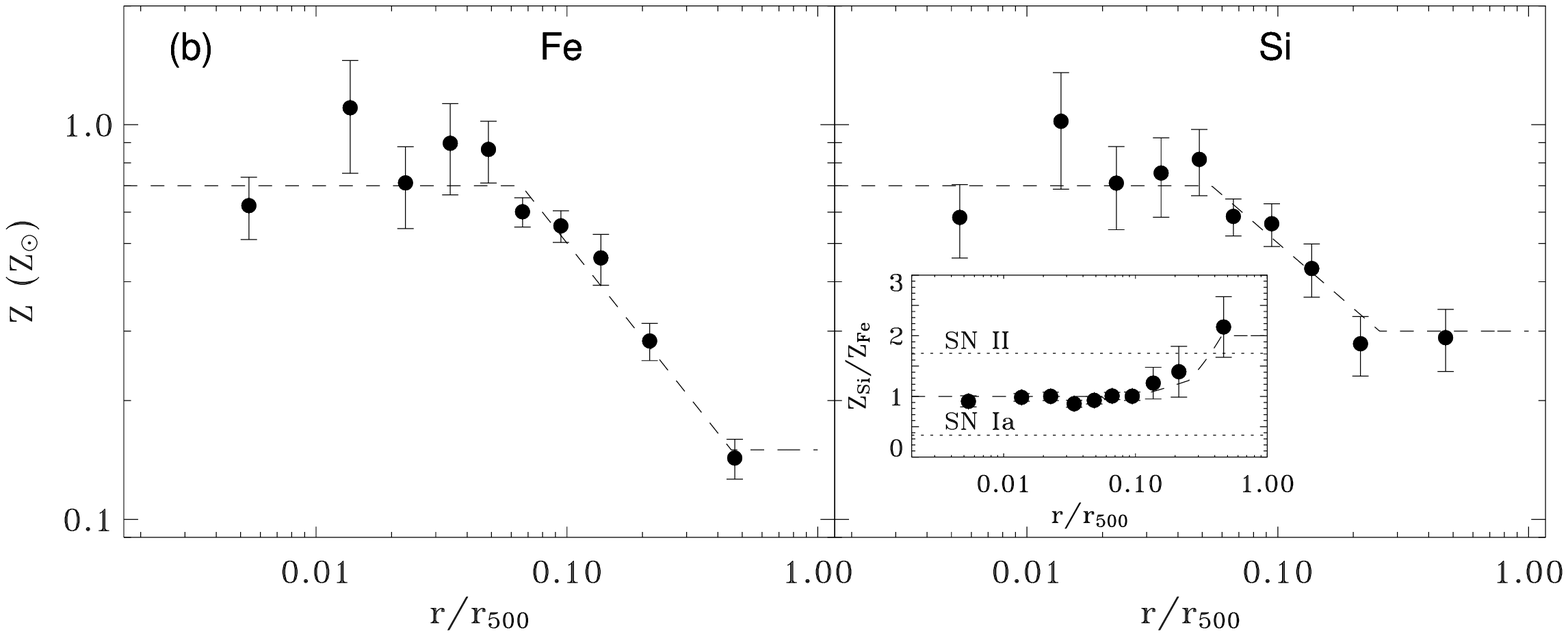}
\caption{({\bf a}) Radial abundance profiles for all groups. Filled circles
  represent data points outside the cool group cores, empty circles
  those within. Solid lines represent the results of regression fits
  to the data outside the group cores. ({\bf b}) The same data, binned into
  radial bins of 20 data points. The inset shows the resulting Si/Fe
  ratio, with dotted lines marking the expectations from pure SN~Ia
  and SN~II enrichment}
\label{rasm:fig1}       
\end{figure}

Profiles of Fe and Si abundance for all groups combined are shown in
Fig.~\ref{rasm:fig1}a. Fe is seen to decline outside the group core,
converging towards a value of $\sim 0.1$\,Z$_\odot$ at $r_{500}$. We
note that this value is lower than that seen in cluster outskirts by a
factor of $\sim 2$ (e.g.\ \cite{rasm:fino00}), even though the Fe
abundance in group cores is comparable to cluster levels.  Si is more
spatially uniform than Fe, declining less steeply outside the core.
Intrinsic scatter is most prominent in the core, where
non-gravitational processes are expected to be important.

These features are all confirmed by Fig.~\ref{rasm:fig1}b, where the
data have been binned into radial bins of 20 measurements, to help
illustrate any radial trends. Adopting the SN model yields of
\cite{rasm:nomo97a,rasm:nomo97b}, the Si/Fe ratios suggest that the
enrichment in group cores can be attributed to a mixture of SN~Ia and
SN~II which is similar to that in the Solar neighbourhood. At large
radii the Si/Fe ratio suggests an increasing predominance of SN~II,
with the abundance pattern being consistent with pure SN~II enrichment
at $r_{500}$.

As the intrinsic scatter in abundances outside the group cores is
fairly small, we made simple parametrizations of the binned profiles
shown in Fig.~\ref{rasm:fig1}b, in order to obtain prescriptions for
$Z_{\rm Fe}(r)$ and $Z_{\rm Si}(r)$ that can be seen as representative
for the entire group sample.  These are shown as dashed lines in
Fig.~\ref{rasm:fig1}b, with the resulting (parametrized) ratio $Z_{\rm
  Si}(r)$/$Z_{\rm Fe}(r)$ conforming to the range allowed by the
adopted SN yields.

\section{Implications}\label{rasm:sec4}

\subsection{Metal Masses and the Role of SN~Ia vs.\ SN~II}\label{rasm:sec4.1}

\begin{figure}
\centering
\includegraphics[height=5.1cm]{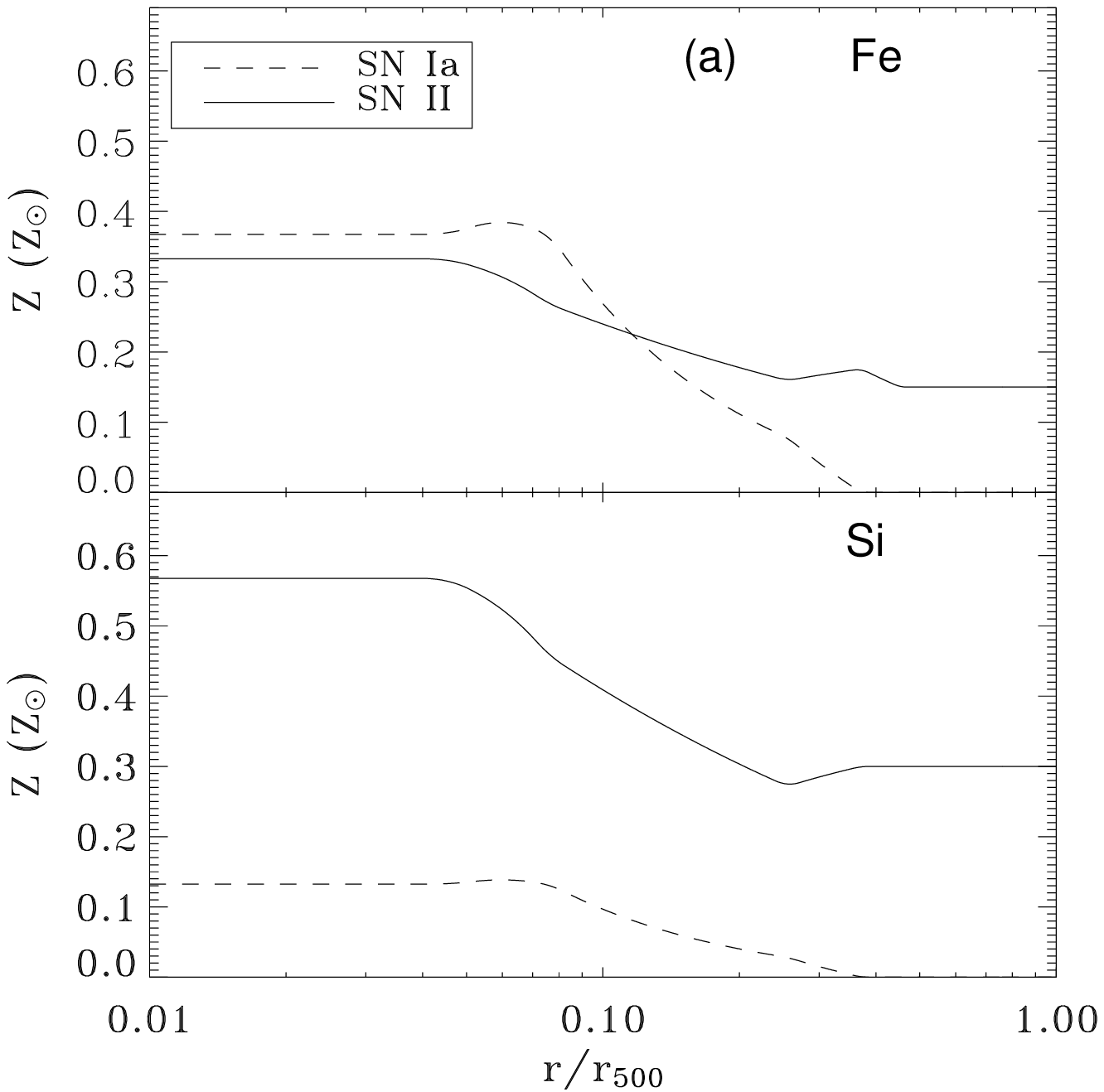}
\includegraphics[height=4.6cm]{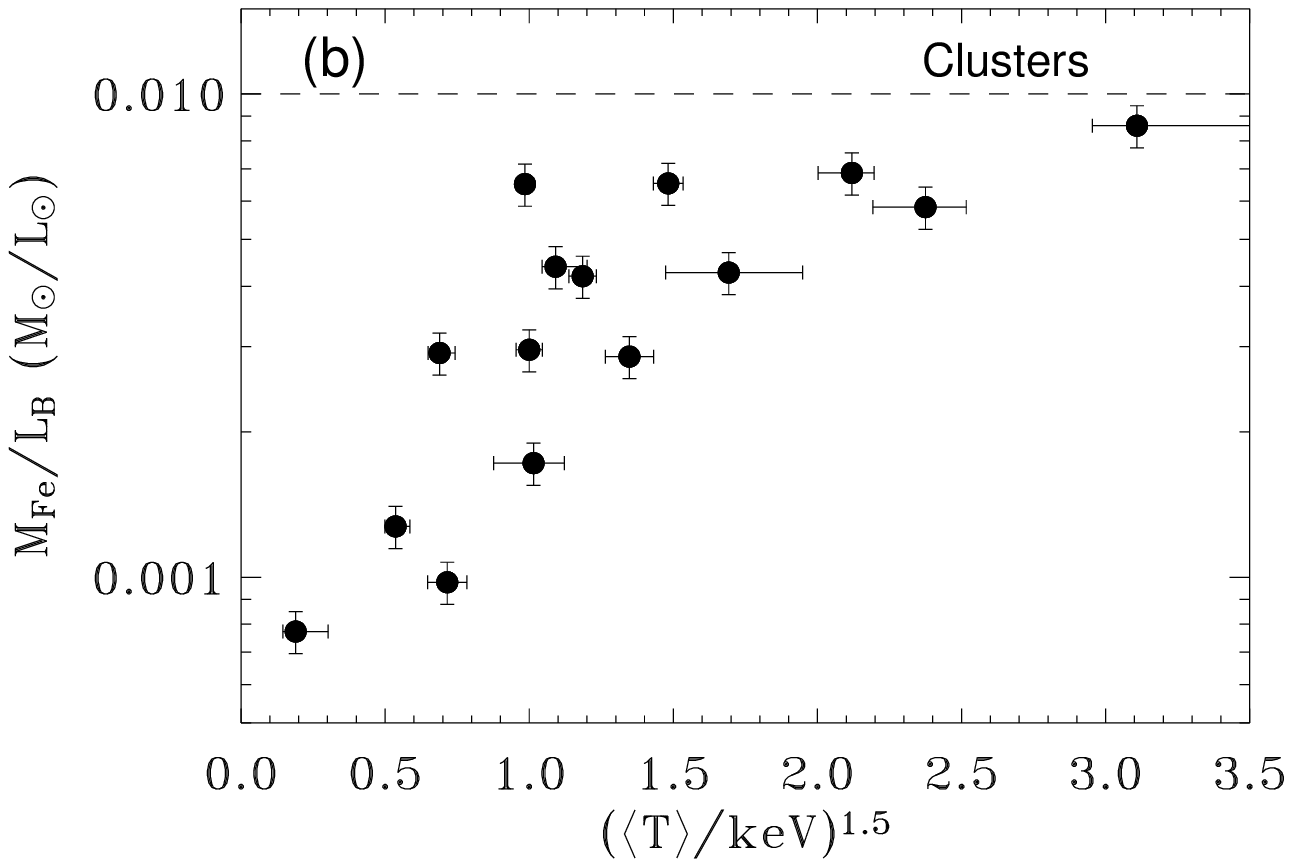}
\caption{({\bf a}) Contribution to the total abundance of Fe and Si
  from SN~Ia and SN~II. ({\bf b}) Fe mass-to-light ratio within
  $r_{500}$ as a function of group 'mass' $M \propto \langle T
  \rangle^{1.5}$.  The dashed line shows the typical cluster value
  \cite{rasm:fino00}}
\label{rasm:fig2}        
\end{figure}

Based on gas density profiles for each group taken from the
literature, the parametrized Fe and Si profiles were decomposed into
contributions from SN~Ia and SN~II.  The result, shown in
Fig.~\ref{rasm:fig2}a, indicates that about half of the iron in group
cores is provided by SN~Ia, but that SN~II products are much more
widely distributed than those of SN~Ia, similarly to results for
clusters.  The total iron mass released by SN within $r_{500}$ spans
the range $3\times 10^{7}$--$3\times 10^{9}$\,M$_\odot$ across the
sample.

Using optical luminosities derived as in \cite{rasm:osmo04}, total
iron mass-to-light ratios within $r_{500}$ are shown in
Fig.~\ref{rasm:fig2}b (results for Si are very similar).  The low Fe
abundance in group outskirts compared to clusters is reflected in
lower Fe $M/L$ ratios.  Even across the fairly narrow range in total
group mass studied here (estimated assuming an empirical
mass-temperature relation $M \propto \langle T \rangle^{1.5}$
\cite{rasm:vikh06}), there is a clear tendency for lower-mass groups
to contain relatively lower amounts of enriched material for their
optical luminosity.

Assuming each SN releases $10^{51}$\,erg directly into the ICM, the
total SN energy range displayed by the groups of $2\times
10^{59}$--$2\times 10^{61}$\,erg is comparable to the total thermal
energy contained in the ICM, with the clear majority of it ($\sim
95$\%) provided by SN~II.

\subsection{The Chemical Enrichment History of Groups}\label{rasm:sec4.2}

Figures~\ref{rasm:fig1} and \ref{rasm:fig2} show that the metals are
not completely mixed throughout the ICM. This applies particularly to
the metals attributed to SN~Ia, of which a considerable fraction is
likely to be associated with prolonged enrichment from the stellar
population of the central galaxy present in all the groups. As also
discussed by \cite{rasm:fino00} on the basis of a small sample of
three groups, the much wider distribution of SN~II products suggests a
scenario of SN~II--dominated enrichment at an early stage in the
formation of the groups, leaving time for the enriched material to mix
well throughout the ICM.

In the group outskirts, the inferred energy per particle imparted to
the IGM by SN~II is of order 0.5\,keV. Coupled with the above, this
indicates that galactic winds associated with an early phase of strong
starburst activity could have contributed substantially to pre-heating
of the ICM in groups.  It is likely that much of this preheating and
the associated chemical enrichment took place before the gas collapsed
into groups, while still located in filaments of low overdensities
\cite{rasm:ponm99}.  Some fraction of the SN ejecta should have been
able to escape from low-mass filaments, giving rise to the trend of
increasing iron mass-to-light ratio with present-day group mass.  The
inferred central rise in SN~II products can then be explained if some
of these metals were released after the group collapsed, possibly
facilitated by galaxy--galaxy and galaxy--ICM interactions, and
potentially with an additional contribution associated with stellar
wind loss from the central elliptical galaxy.

% (i.e.\ there is no need to invoke a burst of recent star formation
% in the central elliptical).

\vspace{3mm}

\noindent{}We are grateful to Stephen Helsdon for his contribution to
the early stages of this work. JR acknowledges the support of an EU
Marie Curie Intra-European Fellowship under contract no.\
MEIF-CT-2005-011171.

\printindex
\end{document}